\documentclass[journal]{IEEEtran}

\usepackage{xcolor,soul,framed} 
\colorlet{shadecolor}{yellow}
\usepackage[pdftex]{graphicx}
\graphicspath{{../pdf/}{../jpeg/}}
\DeclareGraphicsExtensions{.pdf,.jpeg,.png}
\usepackage{cite}
\usepackage[cmex10]{amsmath}
\usepackage{array}
\usepackage{mdwmath}
\usepackage{mdwtab}
\usepackage{eqparbox}
\usepackage{url}
\usepackage{tikz,lipsum,lmodern}
\usepackage{multicol}
\usepackage[most]{tcolorbox}
\begin{document}
\bstctlcite{IEEEexample:BSTcontrol}
\title{ Artificial Neural Network based Modelling for Variational Effect on Double Metal Double Gate Negative Capacitance FET}
\author{Yash~Pathak, Laxman Prasad Goswami ~\IEEEmembership{Student Member,~IEEE,}
        Bansi~Dhar~Malhotra,~\IEEEmembership{ Member,~IEEE,}\\
      
      and~Rishu~Chaujar,~\IEEEmembership{Senior Member,~IEEE}

\thanks{This work was funded  in part by the Council of Scientific and Industrial Research under Grant 08/133(0050)/2020-EMR-I for Yash, and BDM thanks the Science \& amp; Engineering Research Board (SERB), Govt. of India for the honor of a Renowned Fellowship (SB/DF/011/2019).}
\thanks{Y. Pathak and R. Chaujar are with the Department of Applied Physics, Delhi Technological University, Delhi, 110042 India (e-mail: yash$\_$2k19phdap505@dtu.ac.in;  chaujar.rishu@dtu.ac.in).}

\thanks{L. P. Goswami is with the Department of Physics, Indian Institute of Technology, Delhi, 110016 India (e-mail: goswami.laxman@gmail.com).}

\thanks{B. D. Malhotra is with the Department of Biotechnology, Delhi Technological University, Delhi, 110042 India (e-mail: bansi.malhotra@dce.ac.in).}%
 }  

\markboth{IEEE TRANSACTIONS
}{Roberg \MakeLowercase{\textit{et al.}}: High-Efficiency Diode and Transistor Rectifiers}

\maketitle

\begin{abstract}
\boldmath
In this work, we have implemented an accurate machine-learning approach for predicting various key analog and RF parameters of Negative Capacitance Field-Effect Transistors (NCFETs). Visual TCAD simulator and the Python high-level language were employed for the entire simulation process. However, the computational cost was found to be excessively high. The machine learning approach represents a novel method for predicting the effects of different sources on NCFETs while also reducing computational costs. The algorithm of an artificial neural network can effectively predict multi-input to single-output relationships and enhance existing techniques. The analog parameters of Double Metal Double Gate Negative Capacitance FETs (D2GNCFETs) are demonstrated across various temperatures ($T$), oxide thicknesses ($T_{ox}$), substrate thicknesses ($T_{sub}$), and ferroelectric thicknesses ($T_{Fe}$). Notably, at $T=300K$, the switching ratio is higher and the leakage current is $84$ times lower compared to $T=500K$. Similarly, at ferroelectric thicknesses $T_{Fe}=4nm$, the switching ratio improves by $5.4$ times compared to $T_{Fe}=8nm$. Furthermore, at substrate thicknesses $T_{sub}=3nm$, switching ratio increases by $81\%$ from $T_{sub}=7nm$. For oxide thicknesses at $T_{ox}=0.8nm$, the ratio increases by $41\%$ compared to $T_{ox}=0.4nm$. The analysis reveals that $T_{Fe}=4nm$, $T=300K$, $T_{ox}=0.8nm$, and $T_{sub}=3nm$ represent the optimal settings for D2GNCFETs, resulting in significantly improved performance. These findings can inform various applications in nanoelectronic devices and integrated circuit (IC) design.

\end{abstract}



%
\IEEEpeerreviewmaketitle


\section{Introduction}

\IEEEPARstart{S}{ince} 
the 4 decades in the semiconductor industry, the size of FET devices has been decreased to attain better efficiency and performance, in possession with Moore’s law \cite{ref1,ref2,ref3,ref4,ref5}. The new trend of VLSI is governing the variability issue, such as global variability (GV) and local variability (LV) sources for both planar and gate-all-around (GAA) devices. In the case of LV source, a high-k/metal gate device with various work functions, which is vital for statistical variability, would be at extreme risk for device performance because of the random occupied grain orientation of gate material \cite{ref6,ref7}. Additionally, in the case of the GV source, the impact of critical dimension is variability in varying the device performance, and as device size is reduced, this effect cannot be ignored. 

The power and thermal management of integrated circuits (ICs) are vital aspects of the transistor industry. Although with engineering technology and material science advancement transistor generates a low amount of heat. However, the ICs have millions of transistors on the chip, which generate huge amounts of heat. Due to its lesser supply voltage, tunnel FETs have also been cited by many research organizations in the semiconductor world as the ideal FET over traditional MOSFETs for future low-power applications. But dealing with TFET can lead to low Ion, so to get the amalgamation of improved Ion and subthreshold swing, the NC technology is an apt candidate to meet the demand of future electronics. \cite{ref4,ref9,ref10,ref11}. 

However, the variability focuses on the accurate prediction of correlation among different sources. 
Cogenda Visual TCAD and Python are proven tools for doing complicated analysis [3,4]. However, a large number of samples is required for better performance. Hence, a large computational budget is inevitable for precise prediction. In this study, a method of variability analysis is proposed for modeling for device level as well as circuit level with minimum error\cite{ref12,ref13,ref14,ref15,ref16}. Consequently, we propose a new method of machine learning (ML) algorithm with a precise match of the TCAD simulator for improved efficiency. This article consists of an analysis of analog parameters of D2GNCFET for different oxide thicknesses, ferroelectric thickness, temperature, and substrate thickness, which results in reduced subthreshold swing, low input power consumption, reduced leakage current, reduced drain-induced barrier lowering, and better controllability of the device.

\begin{figure}
  \begin{center}
  \includegraphics[width=3.5in]{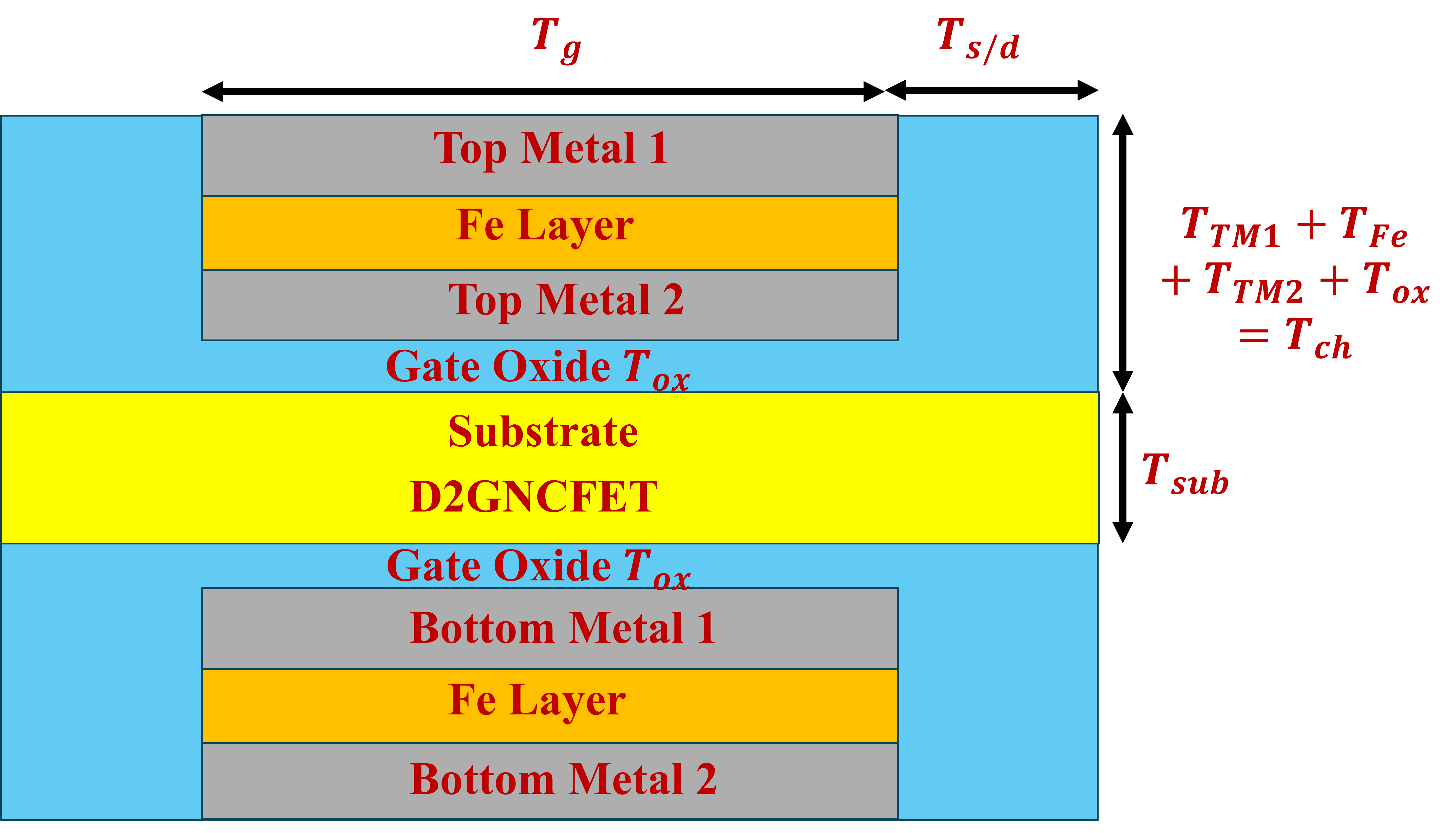}
  \caption{The 2D schematic figure of double metal double gate NCFET. }
\label{fig1}
  \end{center}
\end{figure}


\section{Device Architecture and Methodology}
Fig.\ref{fig1} shows the schematic diagram of the Double metal double gate NCFET (D2GNCFET) \cite{ref17,ref18}. The parameters of D2GNCFET are summarized in table-\ref{tab1}. The channel is made with silicon material and has a length of $L_{g} = 60nm$. The drain and source are made with aluminum which has length $L_{d/s} = 50nm$. The thickness of the substrate ($T_{sub}$), oxide ($T_{ox}$), top metal1/bottom metal1 ($T_{TM1}/T_{TB1}$), top metal2/bottom metal2 ($T_{TM2}/T_{TB2}$) are $5nm$, $0.6nm$, $10nm$, $5nm$ respectively. The total thickness of channel ($T_{ch}$) is $10nm$ as illustrated in table-\ref{tab1}. The material of top metal1/bottom metal1, top metal2/bottom metal2 is Aluminium. The material taken for ferroelectric material, and gate oxide are $HfO_{2}FE$, $SiO_2$. The work function ($\phi$) of aluminum material is fixed at $4.2 eV$ for the double gate. The concentration of the doping profile for Source/Drain ($N_{s/d}$) is $1.0\times 10^{20} cm^{-3}$.

The entire simulation is done by Cogenda Visual TCAD and Python high-level language. The gate voltage ($V_{gs}$) is varied from $0-0.5V$, with drain voltage fixed at $0.05V$. The various physical models are used in this work, such as Shockley Read Hall (SRH) which includes generation and recombination effects, the Arora model which incorporates the effects on carrier mobility due to temperature and impurities concentration, Crowell size for consideration shows the impact ionization, Fermi Dirac statistics is used for improved efficiency and accuracy. 

\begin{table}
\caption{The Device parameters for D2GNCFET  are used for the simulation}
    \begin{center}
        \begin{tabular}{|c|c|p{60pt}|}
        \hline
            Parameter & Symbol & D2GNCFET \\
        \hline
            Substrate thickness & $T_{sub}$ & $5 nm$ \\
            Ferroelectric thickness & $T_{fe}$ & $7 nm$ \\
            Oxide Thickness & $T_{ox}$ & $0.6 nm$ \\
            Gate length & $L_g$ & $50 nm$ \\
        	Concentration of Source/Drain & $N_{s/d}$ & $10^{20} cm^{-3}$ \\
        	Source/Drain length & $L_{s/d}$ & $5 nm$ \\
        	Top/Bottom Metal thickness & $T_{m1}/T_{m1b}$ & $10 nm$ \\
        \hline
        \end{tabular}
    \label{tab1}
    \end{center}
\end{table}

\section{Artificial Neural Network (ANN) Implementation}

\begin{figure*}
  \begin{center}
  \includegraphics[width=6.5in, height = 4in]{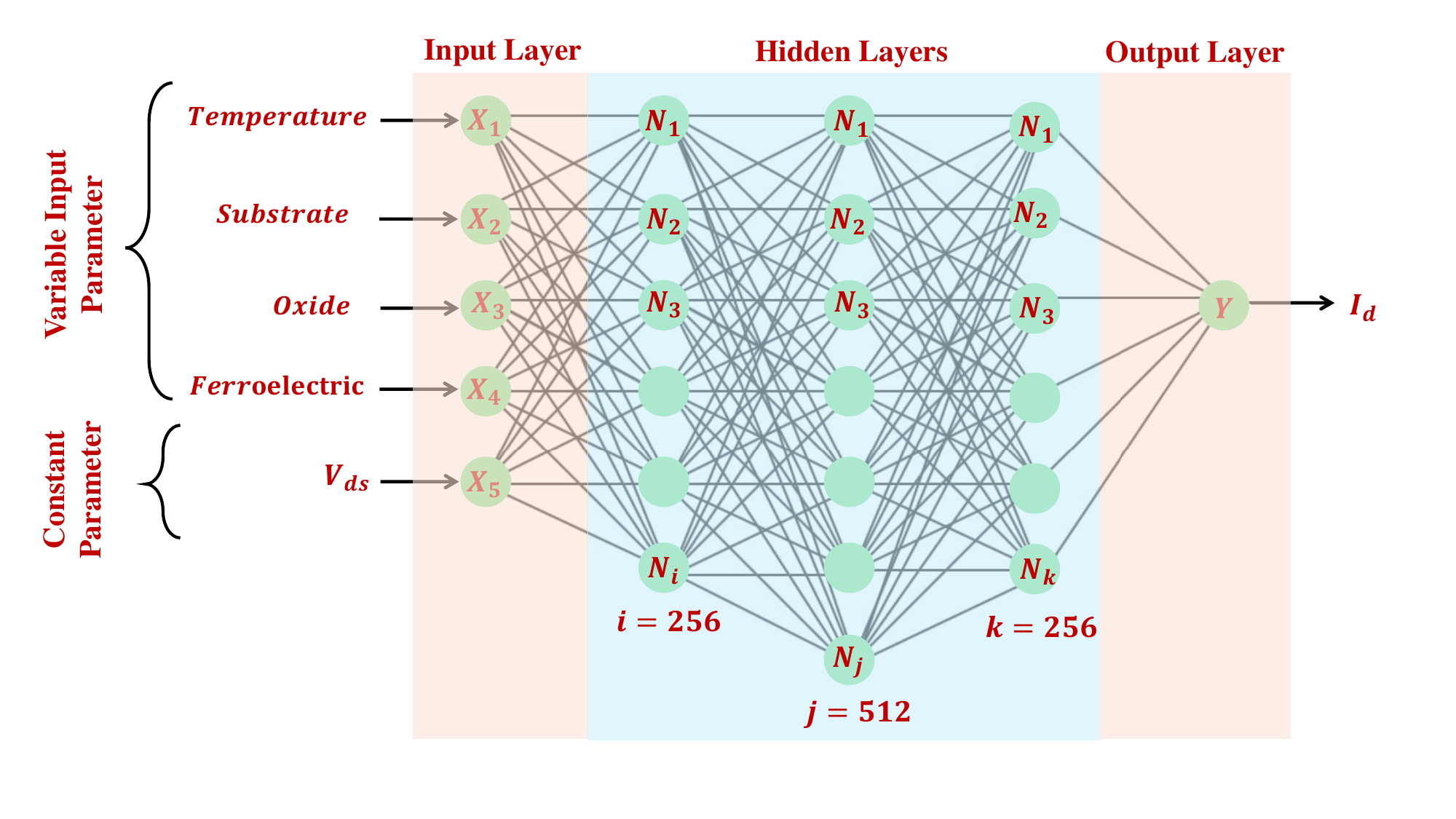}
  \caption{The figure shows the D2GNCFET Artificial Neural Network (ANN) with one input, one output, and three hidden layers. There are $256$ neurons in the first and third hidden layers, and $512$ neurons in the middle layer. All I/O values include the source and the device parameter. The model predicts the drain current based on the input parameters Temperature, Substrate thickness, Oxide thickness, Ferroelectric thickness, and the drain-source voltage $V_{ds}$.}
\label{fig2}
  \end{center}
\end{figure*}

We constructed a tensor flow-based ANN model with a Rectified Linear Unit (ReLU) activation function \cite{hara2015analysis, das2023sers, kessler2017application} for D2GNCFET. The architecture of the ANN model is shown in Fig.\ref{fig2}. The input to the input layer is an array of size 5 for each variable Temperature $T$, Substrate thickness $T_{sub}$, Oxide thickness $T_{ox}$, Ferroelectric thickness $T_{fe}$, and the voltage $V_{ds}$. The output layer provides the ANN prediction for the drain current $I_d$. Further, we consider three hidden layers in our ANN model. The first and third layers have $256$ neurons while there are $512$ neurons in the middle layer. The output of each neuron in the Neural Network (NN) is given as:

\begin{equation}
    a_p^{[q]} = g\left(\Vec{W}_p^{[q]}\cdot\Vec{a}^{[q-1]} + b_p^{[q]}\right)
\end{equation} 

where, $q$ represents the layers of the neural network (NN), and $p$ is the node number such that $a_p^{[q]}$ is the output of $p^{th}$ node from $q^{th}$ layer. $g$ is the activation function. $\Vec{W}$ and $b$ are the weight and biasing parameters for each node. 

\begin{figure}
  \begin{center}
  \includegraphics[width=3.5in, height = 2.5in]{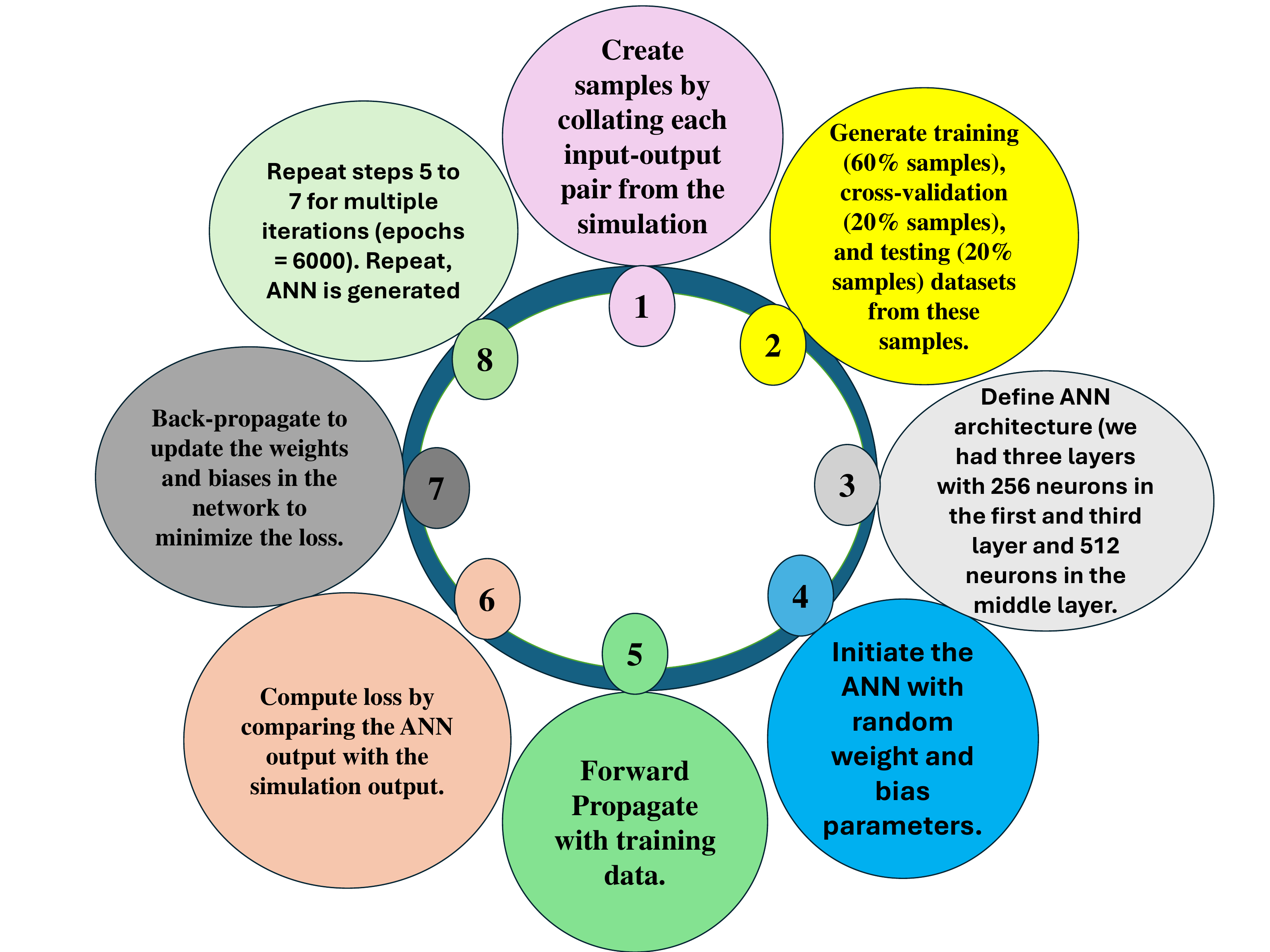}
  \caption{Algorithm for D2GNCFET Artificial Neural Network (ANN) }
\label{flowchart}
  \end{center}
\end{figure}

This function is repeated and runs up to the Taylor series to get the precise result.
The input value put in ANN does not exclude work function variation (WFV) due to WFV impact and varies the $T_{ox}$, $T$, $T_{sub}$, $T_{Fe}$ variability. In addition, it contains fixed parameters to study various dimensions of D2GNCFET. The algorithm of the implemented D2GNCFET ANN is illustrated in fig. \ref{flowchart}.

\begin{figure*}
  \begin{center}
  \includegraphics[width=7.0in, height = 4.5in]{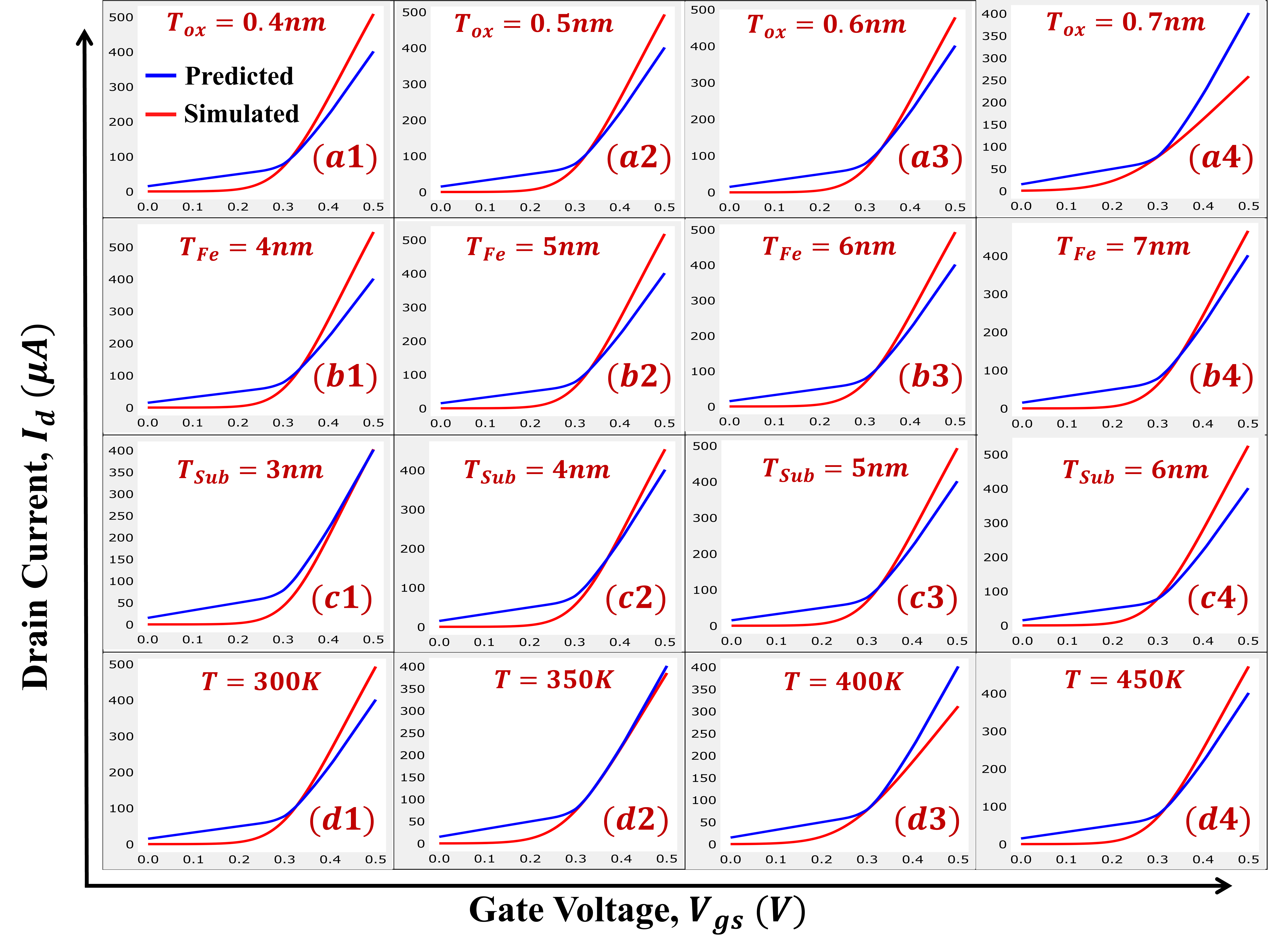}
  \caption{The graph of drain current vs gate voltage for predicted and simulated data of D2GNCFET for different variations of temperature, oxide, thickness, ferroelectric thickness, substrate thickness at V$_d$=0.05V.}
\label{fig3}
  \end{center}
\end{figure*}

\section{Result and discussion}

Fig.\ref{fig3} represents the drain current ($I_{d}$) vs gate voltage ($V_{gs}$) of the TCAD simulation study (in red dash line) and predicted data (blue dash line) by ANN model for D2GNCFET at $V_{ds}=0.05V$. The variation with oxide thickness ($T_{ox}$) are displayed in subplots (a1)-(a4) at $T_{Fe}=7nm$, $T=300K$, $T_{sub}=5nm$. The predicted and simulated data shows a precise match at $T_{ox}=0.6nm$, which reflects the lower error at $0.3V$ of gate voltage. The variation with ferroelectric thickness ($T_{Fe}$) is shown in subplots (b1)-(b4) at $T_{ox}=0.6nm$, $T=300K$, $T_{sub}=5nm$. The precise data of predicted and simulated is matched at $T_{Fe}=8nm$, reflecting the lower error after $0.3V$ of gate voltage.  The variation with substrate thickness ($T_{sub}$) is illustrated in subplots (c1)-(c4) at $T_{ox}=0.6nm$, $T=300K$, $T_{Fe}=7nm$. The precise data of predicted and simulated is matched at $T_{sub}=4nm$, which reflects the lower error at $0.3V$ of gate voltage. The variations with temperature is shown in subplots (d1)-(d4) at $T_{ox}=0.6nm$, $T_{Fe}=7nm$, and $T_{sub}=5nm$. The predicted and simulated data are matched at $T = 300k$, which reflects the minimum error after $0.3V$ of gate voltage.
The different variations of all data variables turned out to coincide at gate voltage at 0.3V. The trainer of predicted data works successfully as a transfer characteristics start from off current to linear current.

\begin{figure*}
  \begin{center}
  \includegraphics[width=6.5in, height = 4in]{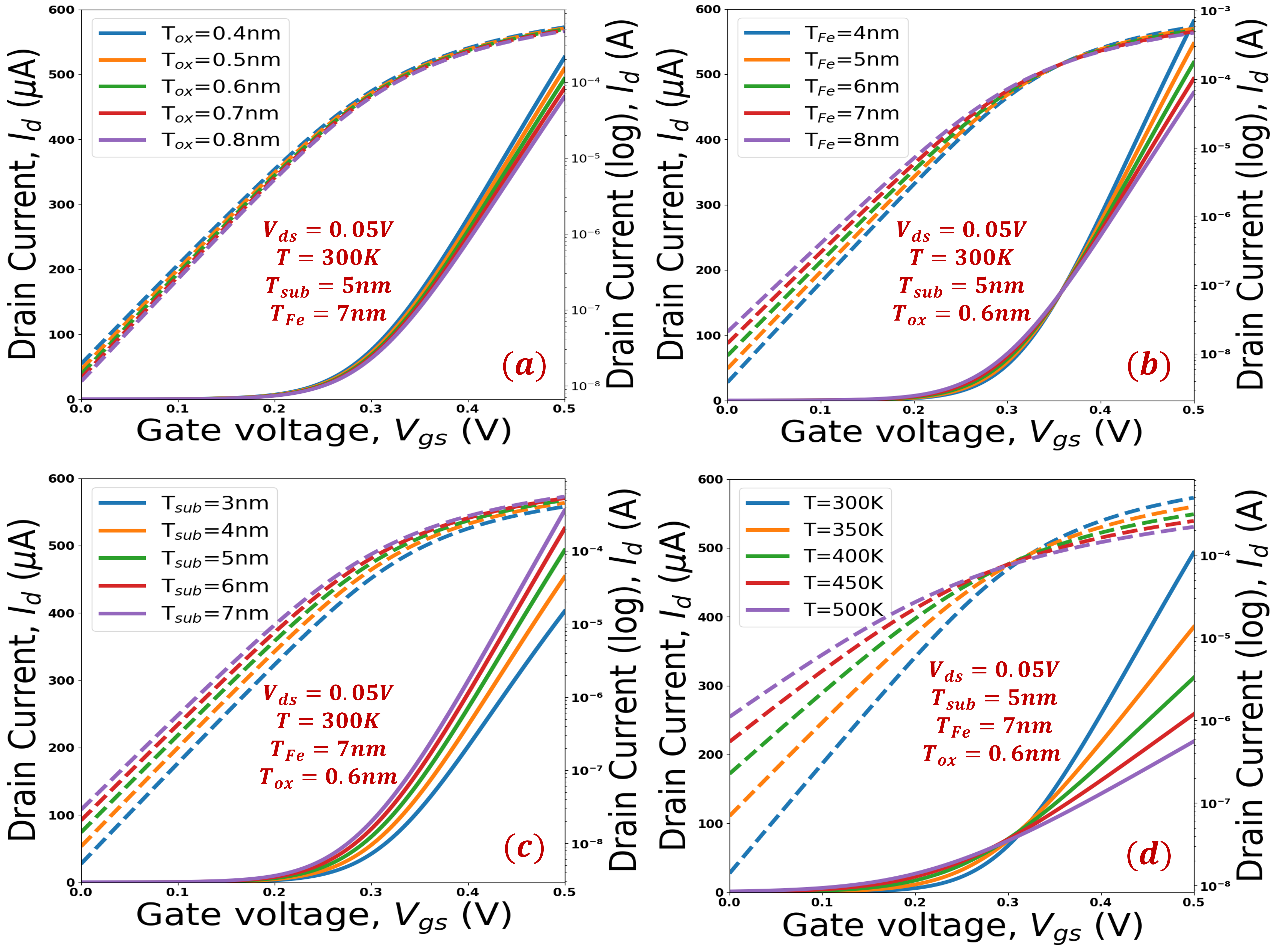}
  \caption{The graph of drain current vs gate voltage D2GNCFET for different variations of temperature, oxide, thickness, ferroelectric thickness, substrate thickness at V$_d$=0.05V.}
\label{fig4}
  \end{center}
\end{figure*}

\begin{figure}
  \begin{center}
  \includegraphics[width=3.5in, height = 3.5in]{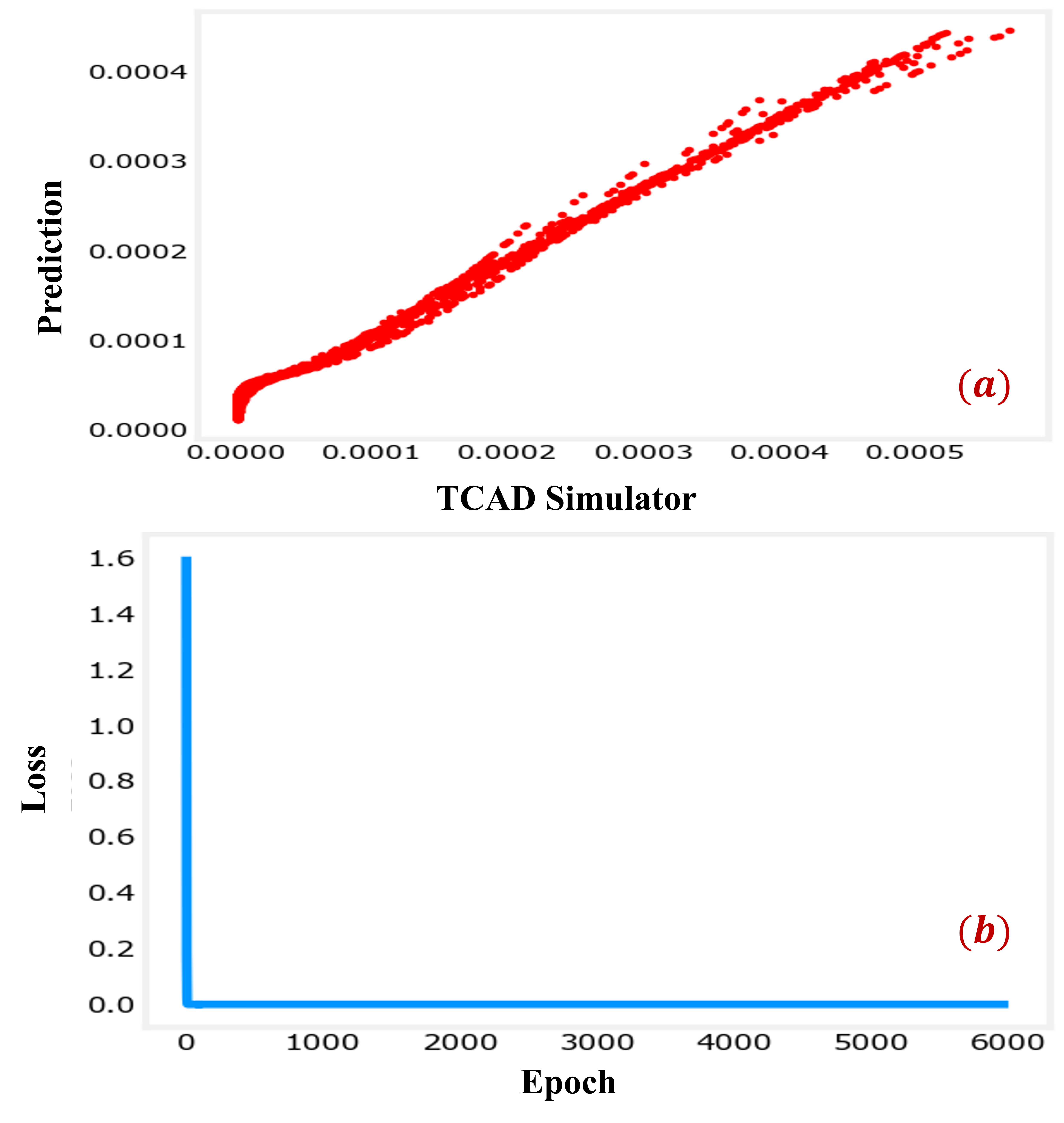}
  \caption{The figure shows the comparison of TCAD simulated output and the output predicted by our ANN model. (a) The graph of predicted versus simulated data and (b) model loss at each epoch during the training of ANN model.}
\label{fig5}
  \end{center}
\end{figure}

\begin{figure*}
  \begin{center}
  \includegraphics[width=6.5in, height = 4in]{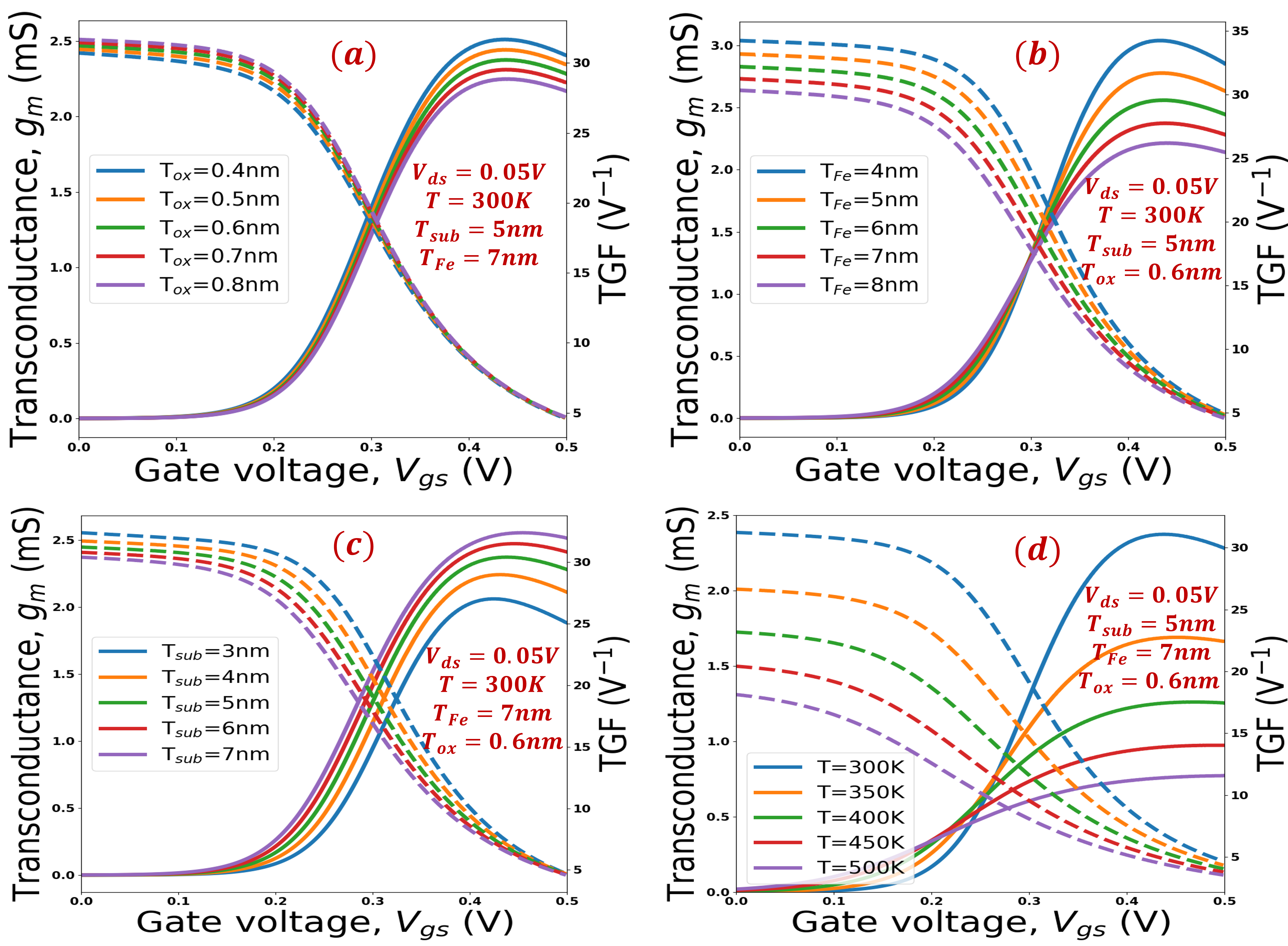}
  \caption{The plot of transconductance and TGF vs V$_{gs}$ for various temperature, thicknesses at V$_d$=0.05V.}
\label{fig6}
  \end{center}
\end{figure*}

\begin{figure*}
  \begin{center}
  \includegraphics[width=6.5in, height = 4in]{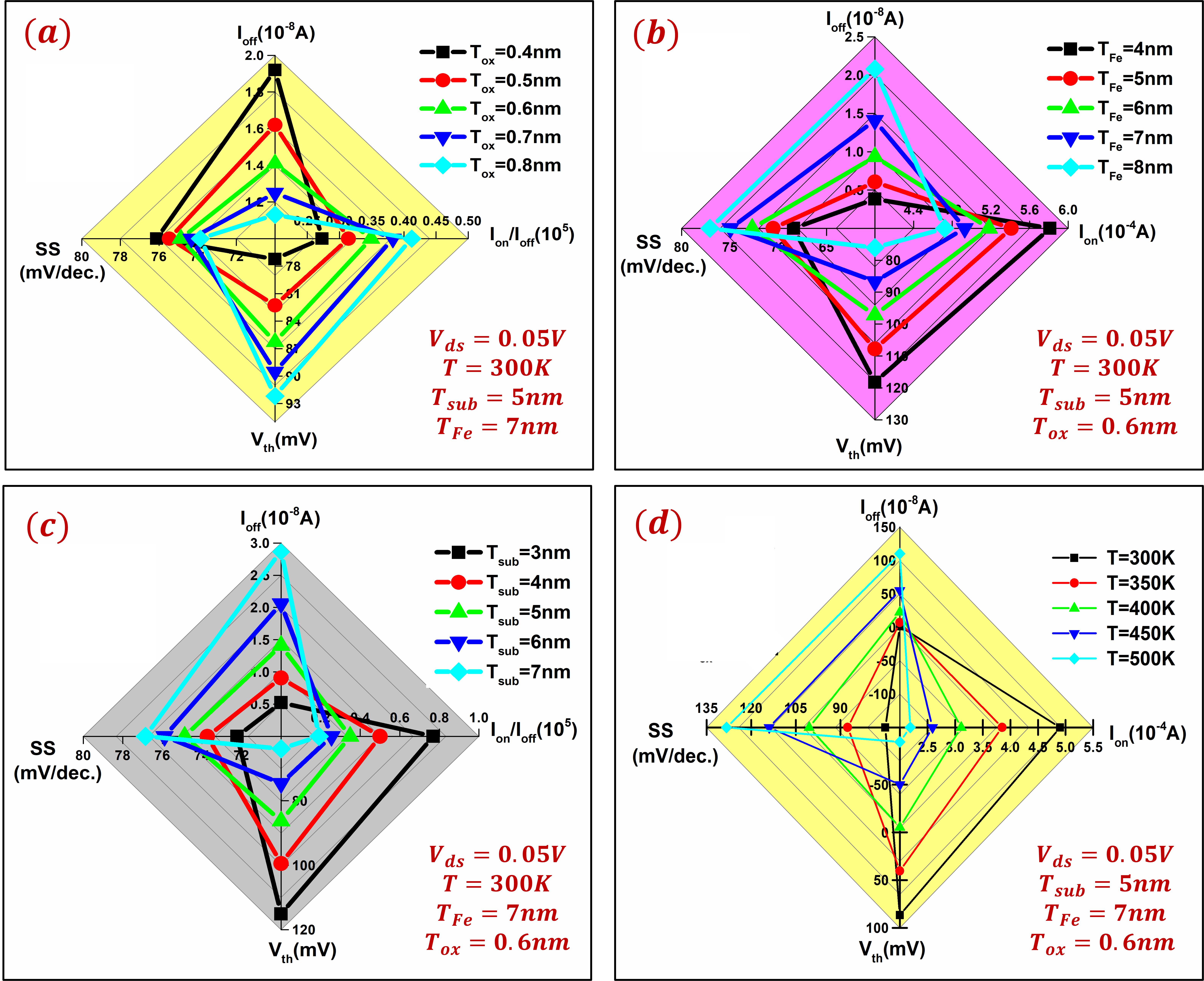}
  \caption{The spider chart of leakage current, On current, Threshold Voltage, SS for different temperature, thicknesses at V$_d$=0.05V.}
\label{fig7}
  \end{center}
\end{figure*}

Fig.\ref{fig4} shows the curve of output characteristics of D2GNCFET in linear and log scales for various temperatures and thicknesses. The device characteristics with variation in oxide thickness($T_{ox}$) are shown in Fig.\ref{fig4}(a) at fixed $V_{ds}=0.05V$, $T=300K$, $T_{Fe}=7nm$, $T_{sub}=5nm$. The better performance of the device is represented at $T_{ox}=0.8nm$, which has a lower leakage current, higher drain current, and better switching ratio, confirming the improved gate controllability. The drain current of D2GNCFET is improved with higher oxide thickness. The impact of variation of ferroelectric thickness($T_{Fe}$) is shown in Fig.\ref{fig4}(b) at fixed $V_{ds}=0.05V$, $T=300K$, $T_{ox}=0.6nm$, $T_{sub}=5nm$. The better performance of the device is represented at $T_{Fe}=4nm$ with lower leakage current and higher drain current, resulting in a better switching ratio with low power consumption. The drain current of D2GNCFET is improved with lesser ferroelectric thickness. The effect of variation of substrate thickness ($T_{sub}$) is shown in Fig.\ref{fig4}(c) at fixed $V_{ds}=0.05V$, $T=300K$, $T_{ox}=0.6nm$, $T_{Fe}=7nm$. The better performance of the device is exhibited at $T_{sub}=3nm$, which has a higher drain current, better switching ratio due to lower leakage current, and improved gate controllability. The drain current of D2GNCFET is improved with lesser substrate thickness. The influence of variation in temperature ($T$) is shown in Fig.\ref{fig4}(d) at fixed $V_{ds}=0.05V$, $T_{Fe}=7nm$, $T_{ox}=0.6nm$, and $T_{sub}=5nm$. The better performance of the device is represented at $T=300K$, which has a lower leakage current and a higher drain current, thus ensuring a better switching ratio, improved gate controllability, and low power consumption. The drain current of D2GNCFET is improved at low temperatures.

The simulation and ANN prediction results are compared in Fig.\ref{fig5}. The linear relation between the prediction and the TCAD simulation results signifies the accuracy of our ANN model as depicted in subplot Fig.\ref{fig5}(a). We further computed the losses during each epoch of training the ANN model. The loss reduces by increasing the number of epochs as shown in subplot Fig.\ref{fig5}(b). 

\begin{table*}
\caption{D2GNCFET comparison for temperature, substrate thickness, ferroelectric, oxide thickness. }
\begin{center}
\begin{tabular}{|c|c|c|c|c|c|}
\hline
 & & & & & \\
No. & Parameter & $I_{off} (10^{-8}A)$ & $I_{on}/I_{off} (10^{5})$ & $V_{th} (mV)$ &	$SS (mV/dec.)$ \\
 & & & & & \\
\hline
 & & & & & \\
$(a1)$ & $T=300k,\ T_{ox}=0.4nm,\ T_{sub}=5nm,\ T_{Fe}=7nm$ & $1.92$ & $0.273$ & $77.23$ & $76.121$ \\
$(a2)$ & $T=300k,\ T_{ox}=0.5nm,\ T_{sub}=5nm,\ T_{Fe}=7nm$ & $1.62$ & $0.314$ & $82.278$ & $75.464$ \\
$(a3)$ & $T=300k,\ T_{ox}=0.7nm,\ T_{sub}=5nm,\ T_{Fe}=7nm$ & $1.25$ & $0.383$ & $89.52$ & $74.33$ \\
$(a4)$ & $T=300k,\ T_{ox}=0.8nm,\ T_{sub}=5nm,\ T_{Fe}=7nm$ & $1.13$ & $0.412$ & $92.18$ & $73.86$ \\
 & & & & & \\
\hline
 & & & & & \\
$(b1)$ & $T=300k,\ T_{ox}=0.6nm,\ T_{sub}=5nm,\ T_{Fe}=4nm$ & $0.382$ & $1.52$ & $118.2$ & $68.4$\\
$(b2)$ & $T=300k,\ T_{ox}=0.6nm,\ T_{sub}=5nm,\ T_{Fe}=5nm$ & $0.605$ & $0.89$ & $107.77$ & $70.53$\\
$(b3)$ & $T=300k,\ T_{ox}=0.6nm,\ T_{sub}=5nm,\ T_{Fe}=6nm$ & $0.935$ & $0.55$ & $97.2$ & $72.69$\\
$(b4)$ & $T=300k,\ T_{ox}=0.6nm,\ T_{sub}=5nm,\ T_{Fe}=8nm$ & $2.08$ & $0.227$ & $76.065$ & $77.061$\\
 & & & & & \\
\hline
 & & & & & \\
$(c1)$ & $T=300k,\ T_{ox}=0.5nm,\ T_{sub}=3nm,\ T_{Fe}=7nm$ & $0.523$ & $0.768$ & $115.107$ & $72.22$ \\
$(c2)$ & $T=300k,\ T_{ox}=0.5nm,\ T_{sub}=4nm,\ T_{Fe}=7nm$ & $0.906$ & $0.5$ & $99.462$ & $73.719$ \\
$(c3)$ & $T=300k,\ T_{ox}=0.5nm,\ T_{sub}=6nm,\ T_{Fe}=7nm$ & $2.06$ & $0.2548$ & $74.58$ & $75.89$ \\
$(c4)$ & $T=300k,\ T_{ox}=0.5nm,\ T_{sub}=7nm,\ T_{Fe}=7nm$ & $2.88$ & $0.192$ & $63.84$ & $76.86$ \\
 & & & & & \\
\hline
 & & & & & \\
$(d1)$ & $T=300k,\ T_{sub}=5nm,\ T_{Fe}=7nm,\ T_{ox}=0.5nm$ & $1.41$ & $0.347$ & $86.68$ & $74.95$ \\
$(d2)$ & $T=350k,\ T_{sub}=5nm,\ T_{Fe}=7nm,\ T_{ox}=0.5nm$ & $6.96$ & $0.0553$ & $40.23$ & $87.55$ \\
$(d3)$ & $T=400k,\ T_{sub}=5nm,\ T_{Fe}=7nm,\ T_{ox}=0.5nm$ & $0.225$ & $0.0138$ & $-5.21$ & $100.498$ \\
$(d4)$ & $T=450k,\ T_{sub}=5nm,\ T_{Fe}=7nm,\ T_{ox}=0.5nm$ & $0.55$ & $0.0047$ & $-50.04$ & $113.927$ \\
$(d5)$ & $T=500k,\ T_{sub}=5nm,\ T_{Fe}=7nm,\ T_{ox}=0.5nm$ & $0.011$ & $0.0012$ & $-94.86$ & $128.28$ \\
 & & & & & \\	    
\hline
\end{tabular}
\label{tab3}
\end{center}
\end{table*}

Fig.\ref{fig6} shows the plot of transconductance and TGF as a function of gate voltage for various temperatures, and different thicknesses at $V_{ds}=0.05V$, $T_{Fe}=7nm$, $T_{ox}=0.6nm$, $T=300K$, $T_{sub}=5nm$. The subplots (a) show the transconductance ($g_{m}$) and transconductance gain factor (TGF) of D2GNCFET at $T_{ox}=0.8nm$ is higher than other oxide thickness, which reflects the high On current, better gain generated per unit cell, reduces the leakage current $V_{ds}=0.05V$, $T_{Fe}=7nm$, $T=300K$, $T_{sub}=5nm$. The subplot (b) shows the transconductance ($g_{m}$) and transconductance gain factor (TGF) of D2GNCFET at $T_{Fe}$=4nm is higher than other thickness of ferroelectric layer, which reveals the high On current, better gain generated per unit cell, reduces the leakage current  V$_{ds}$=0.05V, T=300K, T$_{ox}$=0.6nm, T$_{sub}$=5nm.  The subplot (c) The transcoductance (g$_{m}$) and transconductance gain factor (TGF) of D2GNCFET at T$_{sub}$=3nm is higher than other thickness of substrate, which reflects the high On current, better gain generated per unit cell, reduces the leakage current  V$_{ds}$=0.05V, T$_{Fe}$=7nm, T$_{ox}$=0.6nm, T=300K. The subplot (d) The transcoductance (g$_{m}$) and transconductance gain factor (TGF) of D2GNCFET at T=300K is higher than the other temperatures, which reflects the high On current, better gain generated per unit cell, reduced the leakage current  at V$_{ds}$=0.05V, T$_{Fe}$=7nm, T$_{ox}$=0.6nm, T$_{sub}$=5nm. 

Fig.\ref{fig7} spider chart of leakage current, on current, threshold voltage, SS for different temperatures, and thicknesses of different variable sets. (a) The oxide thickness of 0.8nm shows high On current, lower leakage current, higher threshold voltage, reduced subthreshold swing at V$_{ds}$=0.05V, $T_{Fe}=7nm$, $T_{sub}=5nm$, $T=300K$. (b) The ferroelectric thickness of $4nm$ results in high On current, lower leakage current, higher threshold voltage, reduced subthreshold swing at $V_{ds}=0.05V$, $T_{ox}=0.6nm$, $T_{sub}=5nm$, $T=300K$. (c) The substrate thickness of $3nm$ showcases high On current, lower leakage current, higher threshold voltage, reduced subthreshold swing at $V_{ds}=0.05V$, $T_{Fe}=7nm$, $T_{ox}=0.6nm$, $T=300K$. (d) The temperature of 300K reflects high On current, lower leakage current, higher threshold voltage, reduced subthreshold swing at $V_{ds}=0.05V$, $T_{Fe}=7nm$, $T_{sub}=5nm$, $T_{ox}=0.6nm$. 

\section{Conclusion}

In this study, we have analyzed ANN-based machine learning to access the freedom of variability in the ULSI domain. The Cogenda Visual TCAD simulator and Python high-level language are used throughout the process. The cost of computational work is substantially reduced by ANN-based ML with precise calculation and efficiency. The method investigating variability processes can help in better optimization and designing the useful for ULSI application. The analog parameter of D2GNCFET is studied for various oxide thickness, substrate thickness, temperature, and ferroelectric thickness, such as lower leakage current, higher switching ratio at $T=300K$ by $3000$ times as compared to $T=500K$, at $T_{Fe}=4nm$ by $588\%$ in contrast at $T_{Fe}=8nm$, at $T_{sub}=3nm$ by $300\%$ in comparison at $T_{sub}=7nm$, $T_{ox}=0.8nm$ by $50\%$ from $T_{ox}=0.4nm$. All the analyzed results reveal that $T_{Fe}=4nm$, $T=300K$, $T_{ox}=0.8nm$, $T_{sub}=3nm$ as the most optimum set of variable sources for the D2GNCFET with the most improved performance and can further be used for various applications in nanoelectronic devices and IC designing.

\section*{Acknowledgment}
The authors thanks to Vinod Dham Centre of Excellence for Semiconductors and Microelectronics (VDCE4SM), Nano Bioelectronics and Microelectronics
Lab, Delhi Technological University for giving essential facilities. 



\end{document}